# Adaptive Channel Prediction, Beamforming and Scheduling Design for 5G V2I Network

Tadilo Endeshaw Bogale$^+$, Xianbin Wang$^{++}$ and Long Bao Le$^+$
Institute National de la Recherche Scientifique (INRS)
Université du Québec, Montréal, Canada$^+$
Western University, London, Canada$^{++}$
Email: {tadilo.bogale, long.le}@emt.inrs.ca, xianbin.wang@uwo.ca

*Abstract*—One of the important use-cases of 5G network is the vehicle to infrastructure (V2I) communication which requires accurate understanding about its dynamic propagation environment. As 5G base stations (BSs) tend to have multiple antennas, they will likely employ beamforming to steer their radiation pattern to the desired vehicle equipment (VE). Furthermore, since most wireless standards employ an OFDM system, each VE may use one or more sub-carriers. To this end, this paper proposes a joint design of adaptive channel prediction, beamforming and scheduling for 5G V2I communications. The channel prediction algorithm is designed without the training signal and channel impulse response (CIR) model. In this regard, first we utilize the well known adaptive recursive least squares (RLS) technique for predicting the next block CIR from the past and current block received signals (a block may have one or more OFDM symbols). Then, we jointly design the beamforming and VE scheduling for each sub-carrier to maximize the uplink channel average sum rate by utilizing the predicted CIR. The beamforming problem is formulated as a Rayleigh quotient optimization where its global optimal solution is guaranteed. And, the VE scheduling design is formulated as an integer programming problem which is solved by employing a greedy search. The superiority of the proposed channel prediction and scheduling algorithms over those of the existing ones is demonstrated via numerical simulations.

## I. INTRODUCTION

With the proliferation of mobile internet and smart devices, it is clear that data traffic over wireless networks will continue to grow exponentially in foreseeable future. Driven by many emerging use cases and industry applications, intensive efforts have been undertaken by both academia and industry to design the next generation 5G network. One of the important sectors that uses the 5G network infrastructure is the automotive industry which is currently undergoing dramatic technological transformations, as more and more vehicles are connected to the Internet and with each other [1], [2]. The development of effective vehicular communication systems in highly mobile environment requires accurate understanding about the propagation channel. Furthermore, as 5G base stations (BSs) tend to have large scale antennas, BSs will likely employ beamforming technology to steer their radiation pattern to the desired vehicle equipment (VE). In addition, the current vehicle to vehicle (V2V) communication standard (i.e., IEEE 802.11p) employs a multicarrier system where each of the VEs may use one or more sub-carriers. Thus, efficient and integrated design of channel impulse response (CIR) prediction, beamforming and scheduling is crucial for the successful realization of the future vehicle to infrastructure (V2I) networks.

A number of efforts have been made to estimate the CIR of vehicular communication systems. In [3], least square and minimum mean square error (MMSE) estimation techniques are discussed and their performances in different vehicular Ad-Hoc network (VANET) scenarios are thoroughly investigated. In [4], a cross-layer channel estimator for the 802.11p system where known data are inserted at the higher layers is proposed. The Wiener filter approach to estimate the channel transfer function is discussed in [5]. The filter coefficients are evaluated in IEEE 802.11p systems. In [6], channel estimation and tracking algorithms exploiting the transmitted symbols, time domain truncation, decision-directed feedback, pilot information and V2V channel characteristics is presented. In [2], location aware channel state information (CSI) estimation and beamforming is proposed for millimeter wave vehicular communications. The authors of [7] propose a channel prediction based scheduling strategy for a single antenna system in flat fading channels for data dissemination in VANETs. Along this line, a scheduling method that employs successive interference cancellation (SIC) for V2V systems is discussed in [8].

Measurement results show that the coherence time of vehicular channels is sometimes less than 1 ms due to the high mobility of vehicles in highway scenarios [9]. Furthermore, as the scattering environment can rapidly be changed due to mobility, it is not trivial to find a general CIR statistical model for vehicular channels [7], [10]. These two factors create a significant challenge to learn the instantaneous CIR and its statistical model using the conventional training based CIR estimation approach. This is because the estimated CIR will be likely outdated when it is used to transmit data in the subsequent time slots (this is validated in Section IV). One way of addressing this challenge is by just increasing the training overhead. However, such an approach will dramatically reduce the network throughput especially for a multiuser setup as the training overhead increases with the number of VEs.

For these reasons, the current paper proposes adaptive channel prediction, beamforming and scheduling design that does not utilize the training signal and CIR statistical model. In this regard, first we apply the well known adaptive recursive least squares (RLS) technique (which does not require the channel statistics) for predicting the next block CIR from the past and current block received signals (a block may have one or more orthogonal frequency division multiplexing (OFDM) symbols) [11], [12]. Then, we jointly design the beamforming and VE scheduling for each sub-carrier to maximize the uplink channel average sum rate by utilizing the predicted CIR. The receive beamforming problem is formulated as a Rayleigh quotient problem where its global optimality is guaranteed. And, the VE scheduling design is formulated as a non-convex integer programming problem where we propose the greedy search based algorithm [13]. The superiority of the proposed channel prediction and scheduling algorithms over the existing ones (i.e., algorithms that employ outdated CIR with a greedy scheduling and predicted CIR with a random VE selection) is demonstrated by numerical simulations.

This paper is organized as follows. Section II discusses the system and channel models. In Section III, the proposed CIR prediction, and joint beamforming and scheduling designs are presented. Computer simulation results are provided in Section IV. Finally, conclusions are drawn in Section V.

## II. SYSTEM AND CHANNEL MODEL

We consider a multiuser system with a transmission scheme having a sampling period $T_s$ and a channel having a maximum delay spread $T_d$. It is assumed that the BS and each VE are equipped with $N$ antennas and 1 antenna, respectively. For these settings, the number of multipath CIR taps $L$ is approximated as $L = \frac{T_d}{T_s}$ [14]. The multipath coefficients between the $k$th VE to the $n$th BS antenna is denoted as [15]

$$\bar{\mathbf{h}}_{kn} = [\bar{h}_{kn1}, \bar{h}_{kn2}, \cdots, \bar{h}_{knL}]^T. \quad (1)$$

The analytical model for $\bar{\mathbf{H}}_k = [\bar{\mathbf{h}}_{k1}, \bar{\mathbf{h}}_{k2}, \cdots, \bar{\mathbf{h}}_{kN}]$ may vary from one standard to another. In this paper, we assume that each element of $\bar{\mathbf{h}}_{kn}$ is independent of others. However, $\bar{h}_{knl}$ is correlated over time. We further assume that the spatial correlation matrix of all multipath components are the same. Under these assumptions, $\bar{\mathbf{H}}_k$ can be expressed as [16]

$$\bar{\mathbf{H}}_k = \sqrt{\mathbf{P}_k}\tilde{\mathbf{H}}_k\sqrt{\mathbf{R}_k} \quad (2)$$

where each entry of $\tilde{\mathbf{H}}_k$ is an independent and identically distributed (i.i.d) zero mean circularly symmetric complex Gaussian (ZMCSCG) random variable with unit variance, $\mathbf{R}_k \in \mathcal{C}^{N \times N}$ is a positive semidefinite spatial covariance matrix, and $\mathbf{P}_k = \text{diag}(g_{k1}^2, g_{k1}^2, \cdots, g_{kL}^2) \in \Re^{L \times L}$ with $g_{kl}^2$ being the channel gain corresponding to the $l$th path.

In most wireless standards (e.g., 802.11p), OFDM based transmission is commonly adopted. For such a transmission, the channel coefficient of each sub-carrier has practical importance. In fact, this coefficient can be obtained by linearly combining the multipath tap coefficients (1). To this end, the channel between the $k$th VE to the BS's $n$th antenna in sub-carrier $s$ can be expressed as [14], [17]

$$h_{kns} = \mathbf{f}_s^H \bar{\mathbf{h}}_{kn} \quad (3)$$

where $\mathbf{f}_s^H = [1, e^{-j\frac{2\pi}{M}s}, e^{-j\frac{2\pi}{M}2s}, \cdots, e^{-j\frac{2\pi}{M}(L-1)s}]$ with $M$ being the fast Fourier transform (FFT) size of the OFDM. Under this model, one can consider the CIR prediction, beamforming and scheduling designs either for the uplink or downlink channels. Here, we consider the uplink channel, and the $s$th sub-carrier $k$th VE signal to interference plus noise ratio (SINR) for this channel can be expressed as[1]

$$\gamma_{ks} = \frac{P_{ks}|\mathbf{w}_{ks}^H \mathbf{h}_{ks}|^2}{\sum_{i \neq k}^K P_{is}|\mathbf{w}_{ks}^H \mathbf{h}_{is}|^2 + \sigma^2|\mathbf{w}_{ks}|^2} \quad (4)$$

where $\mathbf{h}_{ks} = [h_{k1s}, \cdots, h_{kNs}]^T$, $P_{ks}$ ($\mathbf{w}_{ks}$) is the $k$th VE $s$th sub-carrier transmitted power (receive beamforming vector) and $\sigma^2$ is the noise power at the BS. We assume that $\bar{\mathbf{H}}_k$ is constant at least for $S_c$ OFDM symbols where $S_c$ is related to the channel coherence time[2]. The $i$th block CIR for the $k$th VE $\bar{\mathbf{H}}_k[i]$ can thus be given as [18]

$$\bar{\mathbf{H}}_k[i] = \sqrt{\mathbf{P}_k[i]}\tilde{\mathbf{H}}_k[i]\sqrt{\mathbf{R}_k[i]}. \quad (5)$$

## III. CIR PREDICTION, BEAMFORMING AND SCHEDULING DESIGN

This section discusses the proposed CIR prediction, joint beamforming and scheduling design algorithms.

### A. CIR prediction design

As stated in the introduction section, the paper assumes that the CIR statistical model is not known. Thus, the proposed approach may need to jointly predict $\sqrt{\mathbf{P}_k[i+1]}$, $\sqrt{\mathbf{R}_k[i+1]}$ and $\tilde{\mathbf{H}}_k[i+1]$ from the past and current block received signals. Since the time scale at which $\sqrt{\mathbf{P}_k[i]}$ and $\sqrt{\mathbf{R}_k[i]}$ are varied is much larger than that of $\tilde{\mathbf{H}}_k[i]$ [19], [20], we predict $\tilde{\mathbf{H}}_k[i+1]$ by assuming that $\sqrt{\mathbf{P}_k[j]}$ and $\sqrt{\mathbf{R}_k[j]}$ are constant for $j = i+1, i, \cdots, i-C_b+1$, where $\sqrt{\mathbf{P}_k[i]}$ and $\sqrt{\mathbf{R}_k[i]}$ are computed from the $C_b$ most recent block received signals and $C_b$ is the design parameter. In the following, we discuss the proposed algorithm to predict $\tilde{\mathbf{H}}_k[i+1]$.

For better mathematical tractability, we denote the Eigenvalue decomposition of $\mathbf{R}_k$ as (we ignore here the time index i since it is clear from the context)

$$\text{svd}(\mathbf{R}_k) \triangleq \mathbf{U}_k \mathbf{D}_k \mathbf{U}_k^H \quad (6)$$

where $\mathbf{D}_k$ is a $Q_k \times Q_k$ sized diagonal matrix containing the non-zero eigenvalues of $\mathbf{R}_k$ arranged in decreasing order. As $Q_k \leq N$, we can represent the low dimensional equivalent CIR matrix of $\tilde{\mathbf{H}}_k \in \mathcal{C}^{L \times Q_k}$ using (6) as

$$\tilde{\tilde{\mathbf{H}}}_k = \sqrt{\mathbf{P}_k^{-1}} \bar{\mathbf{H}}_k \mathbf{U}_k \sqrt{\mathbf{D}_k^{-1}}. \quad (7)$$

Since each element of $\tilde{\tilde{\mathbf{H}}}_k[i]$ is independent, the CIR prediction can be done for each entry of $\tilde{\tilde{\mathbf{H}}}_k[i+1]$ separately. We utilize the RLS algorithm to predict $\tilde{\tilde{\mathbf{H}}}_k[i+1]$ from $\tilde{\tilde{\mathbf{H}}}_k[i], \cdots, \tilde{\tilde{\mathbf{H}}}_k[i-S_b+1]$ where $S_b$ is the design parameter. The RLS algorithm is explained as follows [12], [21].

Denote an element of $\tilde{\tilde{\mathbf{H}}}_k[i]$ as $f[i]$ (i.e., ignoring $k$ and the entry indexes). The main idea of the RLS algorithm is to predict $f[i+1]$ from $\mathbf{f}[i] \triangleq [f[i], \cdots, f[i-S_b+1]]^T$ as

$$\hat{f}[i+1] = \mathbf{q}[i]^T \mathbf{f}[i]. \quad (8)$$

---

[1]The methodology used in this paper can be applied for the downlink channel.

[2]In the following, we use the word block to denote $S_c$ consecutive OFDM symbols.

And the predictor $\mathbf{q}[n]$ is calculated to optimize

$$\min_{\mathbf{q}[n]} \sum_{j=n-S_b+1}^{n} \lambda^{n-j} |f[j+1] - \mathbf{q}[n]^T \mathbf{f}[j]|^2 \quad (9)$$

where $0 < \lambda \leq 1$ is a forgetting factor that accounts for a possible non stationarity of the input samples $\mathbf{f}[j]$. After doing some straightforward steps, one can compute $\mathbf{q}[n]$ recursively as [12], [21]

$$\mathbf{q}[n] = \mathbf{q}[n-1] + \boldsymbol{\tau}[n-1]e[n] \quad (10)$$

where $e[n] = f[n] - \hat{f}[n]$,

$$\boldsymbol{\tau}[n] = \frac{\mathbf{Z}[n-1]\mathbf{f}[n]}{\lambda + \mathbf{f}[n]^H \mathbf{Z}[n-1]\mathbf{f}[n]} \quad (11)$$

and the matrix $\mathbf{Z}$ is the inverse of the sample covariance matrix $\sum_j \lambda^{n-j} \mathbf{f}[j]\mathbf{f}[j]^H$ which can be calculated recursively as

$$\mathbf{Z}[n] = \frac{1}{\lambda}(\mathbf{I} - \boldsymbol{\tau}[n]\mathbf{f}[n]^H)\mathbf{Z}[n-1]. \quad (12)$$

One can observe from (10) that $\mathbf{q}[n]$ depends on $\mathbf{q}[n-1]$ and $\boldsymbol{\tau}[n-1]$. This recursive behavior requires appropriate parameter initializations. In the current paper, we employ $\lambda = 0.999$, $\mathbf{Z}[0] = \delta \mathbf{I}$ and $\mathbf{q}[0] = \mathbf{0}$ with $\delta = 100$ [12][3].

Once all the elements of $\tilde{\bar{\mathbf{H}}}_k[i+1]$ are predicted, the next task is to perform the joint beamforming and scheduling design which is presented in the next subsection.

### B. Joint beamforming and VE scheduling design

Once the CIR of each VE is predicted, the next task is to schedule VEs to utilize one or more sub-carriers. In practice, a scheduler is designed to optimize some performance criteria. In the current paper, we consider the maximization of the total sum rate achieved at each sub-carrier corresponding to the first OFDM symbol of the $(i+1)$th block under a per VE power constraint which can be mathematically formulated as[4]

$$\max_{K_s, \mathbb{K}_s, \mathbf{w}_{ks}} \sum_{k=1}^{K_s} \log(1 + \gamma_{ks}^p), \quad P_{ks} \leq P_{max}, \quad k \in \mathbb{K}_s \quad (13)$$

where $\gamma_{ks}^p(\mathbf{w}_{ks})$ is the $k$th VE $s$th sub-carrier SINR (receive beamforming vector), $P_{max}$ is the maximum transmitted power at each sub-carrier of a VE and $\mathbb{K}_s$ is the scheduled set of VEs at sub-carrier $s$. Note that since $\gamma_{ks}^p$ is computed using the predicted CIR, this SINR could be different from the true $\gamma_{ks}$ achieved at the first OFDM symbol of the $(i+1)$th block.

The VE scheduling problem (13) is an integer programming non-convex problem which prohibits us from getting the global optimal solution. Here we propose to solve the above problem by utilizing greedy scheduling technique where we sequentially increase the number of VEs (i.e., $K_s$) while ensuring a non-decreasing sum rate [13]. In fact, for fixed $K_s$, the global

---

[3]These initializations may not necessarily yield the optimal performance. More information about the optimal selection of $\lambda$ can be found in [22].

[4]Note that some applications require a minimum $\gamma_{ks}^p$ to ensure a certain service quality. And considering the rate maximization problem by taking into account the minimum $\gamma_{ks}^p$ is left for future research.

---

optimal solution that maximizes (4) (which also maximizes (13)) can be obtained by applying the generalized Rayleigh quotient approach [23]. The proposed joint beamforming and VE scheduling algorithm is summarized in **Algorithm I**.

**Algorithm I**: Joint beamforming and scheduling algorithm.
**Input**: The number of VEs to be scheduled ($K_t$).
   **Initialization**: Set $\mathbb{K}_t = \{1, 2, \cdots, K_t\}$, $\mathbb{K}_s = \emptyset$, $K_s = 0$ and $f_s(\mathbf{W}_s) = 0$, where $f_s(\mathbf{W}_s)$ is the objective function of (13) with $\mathbb{K}_s$ VEs and $\mathbf{W}_s = [\mathbf{w}_{1s}, \mathbf{w}_{2s}, \cdots, \mathbf{w}_{K_s s}]$.
   **while** $K_s \leq K_t$ **do**
      • Set $u = \emptyset$.
      **for** $j \in \mathbb{K}_t$ and $j \notin \mathbb{K}_s$ **do**
         • Compute $f_s(\tilde{\mathbf{W}}_s)$ with $\tilde{\mathbb{K}}_s = \mathbb{K}_s \cup \{j\}$ VEs, where $\tilde{\mathbf{W}}_s$ is the receive beamforming for the VEs in $\tilde{\mathbb{K}}_s$.
         **if** $f_s(\tilde{\mathbf{W}}_s) > f_s(\mathbf{W}_s)$ **then**
            • Set $u = j$ and $f_s(\mathbf{W}_s) = f_s(\tilde{\mathbf{W}}_s)$.
         **end if**
      **end for**
      **if** $u = \emptyset$ **then**
         • **Break**.
      **else**
         • Set $\mathbb{K}_s = \mathbb{K}_s \cup \{u\}$ and $K_s = K_s + 1$.
      **end if**
   **end while**

**Practical Issues**: Up to now, we have provided the framework for the joint adaptive channel prediction, beamforming and scheduling algorithms. In fact, these algorithms require the predicted CIR which is computed by employing $\bar{\mathbf{H}}_k[j], j \leq i, \forall k$. However, since more than one VEs could be scheduled to use a given sub-carrier, $\bar{\mathbf{H}}_k[j]$ may be observed with additional noise and the CIR of other VEs. In a general setup, the received signal at the $s$th sub-carrier and $j$th OFDM symbol can be expressed as

$$\mathbf{y}_s[j] = \mathbf{f}_s^H \bar{\mathbf{H}}_k[j] x_{ks}[j] + \sum_{i=1}^{K_s-1} \mathbf{f}_s^H \bar{\mathbf{H}}_i[j] x_{is}[j] + \mathbf{n}_s[j] \quad (14)$$

where $x_{ks}[j]$ is the transmitted signal from the $k$th VE at the $s$th sub-carrier of $j$th OFDM symbol, and $\mathbf{n}_s[.]$ is the received noise at the $s$th sub-carrier where each of its entries is i.i.d ZMCSCG random variable with variance $\sigma^2$.

One can examine (14) for two cases: $K_s = 1$ and $K_s > 1$. In the case of $K_s = 1$, if the $k$th VE utilizes $L$ sub-carriers $\mathbf{F}_L = [\mathbf{f}_{s_1}, \mathbf{f}_{s_2}, \cdots, \mathbf{f}_{s_L}]$ that ensures $\mathbf{F}_L \mathbf{F}_L^H \approx \mathbf{I}$ and $x_{ks}, \forall s$ are known perfectly, $\bar{\mathbf{H}}_k[j]$ can be estimated as [15]

$$\hat{\bar{\mathbf{H}}}_k[j] = \bar{\mathbf{H}}_k[j] + \bar{\mathbf{n}}[j] \quad (15)$$

where the entries of $\bar{\mathbf{n}}[.]$ are i.i.d ZMCSCG random variables each with variance $\sigma^2$. When $K_s > 1$, one can achieve the CIR expression similar to (15) if there exists a set of sub-carriers $\mathbb{O}_k$ such that $x_{ks}, s \in \mathbb{O}_k$ is orthogonal to $x_{is}, s \in \mathbb{O}_k, \forall i \neq k$.

In the current paper, we assume that the transmitted symbols $x_{ks}[j], \forall k, j$ are recovered perfectly at the BS from the received signal $\mathbf{y}_s[j]$ (i.e., the predicted CSI corresponding to the $j$th OFDM symbol has sufficient quality to recover $x_{ks}[j], \forall k, j$ without any error) and the number of OFDM symbols per block $S_c$ is selected such that the observed samples corresponding to $\bar{\mathbf{H}}_k[j]$ is as given in (15) for all VEs. In addition, the spatial covariance matrix $\mathbf{R}_k$ of (6) is computed by taking into account the effect of $\bar{\mathbf{n}}[.]$ like in [24].

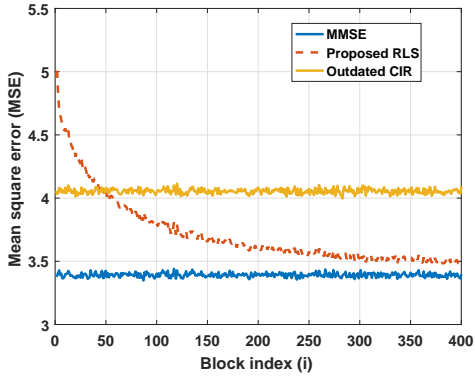

Fig. 1. Comparison of MSEs achieved by different CIR predictors.

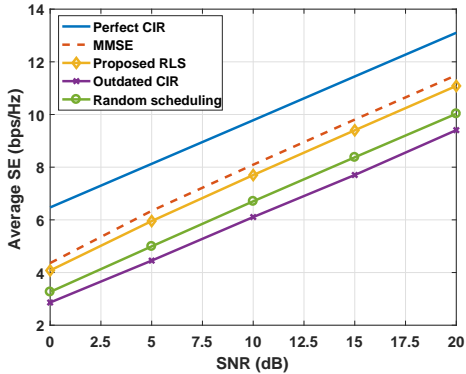

Fig. 2. Average SE achieved with single user scheduling.

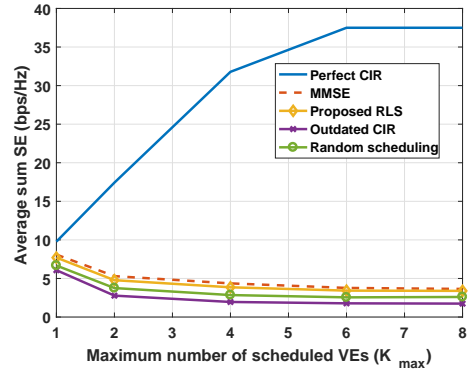

Fig. 3. Effect of the maximum number of scheduled VEs $K_{max}$.

Thus, in practice, the proposed algorithms require training signals just at the beginning of the transmission duration or when the BS is not able to reliably decode the transmitted signals $x_{ks}[j]$ which occurs quiet rarely.

## IV. SIMULATION RESULTS

This section presents simulation results. We consider a typical scenario of IEEE 802.11p standard where the relative powers of $g_{kl}, \forall k, l$ are [0 dB, $-6.3$ dB, $-25.1$ dB, $-22.7$ dB] (i.e., VTV - Expressway Oncoming without Wall, 300 m - 400 m scenario) [25], [26]. The spatial channel covariance matrices for this environment is characterized for the commonly adopted uniform linear array (ULA) model as

$$\mathbf{R}_k = \sum_{i=1}^{C_k} \mathbf{a}(\theta_{ki})\mathbf{a}(\theta_{ki})^H \qquad (16)$$

where $C_k$ is the number of clusters in the environment corresponding to the $k$th VE CIR, $\theta_{ki}$ is the angle of arrival (AoA) for the $i$th cluster, and $\mathbf{a}(\theta) = \frac{1}{\sqrt{N}}[1, e^{j\frac{2\pi}{\tilde{\lambda}}d\sin(\theta)}, \cdots, e^{j(N-1)\frac{2\pi}{\tilde{\lambda}}d\sin(\theta)}]^T$ with $\tilde{\lambda}$ as the transmission wave length and $d = 0.5\tilde{\lambda}$ is the antenna spacing [24]. We assume the well known Jake's model to capture the channel correlation between $\tilde{\mathbf{H}}_k[i]$ and $\tilde{\mathbf{H}}_k[j]$ as

$$E\{(\tilde{\mathbf{H}}_k[i])_{m,n}(\tilde{\mathbf{H}}_k[j])^H_{m,n}\} = \alpha_{kij} \qquad (17)$$

where $\alpha_{kij} = J_0(2\pi f_k|i-j|)$ with $J_0(.)$ as the zeroth-order Bessel function and $f_k$ is the maximum Doppler frequency normalized by the sampling rate $\frac{1}{T_s}$ [12].

Furthermore, we set $L = 4$, $M = 64$, $N = 8$, $T_s = 166\mu s$, $C_b = S_b = 8$, the carrier frequency as 5.6 GHz, the speed of each VE as 90 km/hr, $\mathbf{R}_k = \mathbf{R}, \forall k$ with $C_k = 8$ and $\theta_{ki}, \forall i$ are selected randomly. The signal to noise ratio (SNR) for each sub-carrier is defined as $\gamma = \frac{P_{max}}{\sigma^2}$ and is controlled by varying $P_{max}$ while keeping $\sigma^2$ constant. All of the plots are generated for $s = 4$ by averaging 10000 channel realizations[5].

### A. CIR prediction

This simulation discusses the accuracy of the predicted CIR coefficients and convergence of the proposed RLS algorithm when $\gamma$ is set to $\gamma = 20$ dB. We compare the mean square error (MSE) values between the true and predicted CIR coefficients for three algorithms in Fig. 1: the MMSE predictor which requires the CIR statistics, the proposed RLS predictor and the design that utilizes outdated CIR. This figure shows that employing the outdated CIR suffers from high prediction error. Furthermore, the MSE achieved by the proposed RLS approach decreases with $i$ and achieves closer to that of the MMSE approach in around 200 blocks. This confirms that the proposed RLS predictor converges to the optimal estimator[6].

### B. Joint beamforming and scheduling

This subsection presents the performance of the proposed joint beamforming and scheduling design. In this regard, we set the total number of VEs as $K_t = 24$ and $K_s \leq K_{max}$. We compare the average spectrum efficiency (SE) (i.e., average rate per Hertz) achieved by five approaches as shown in Fig. 2 when $K_{max} = 1$: perfect CIR, MMSE predicted CIR, proposed RLS predicted CIR and outdated CIR, all with the proposed scheduling, and proposed RLS predicted CIR with random scheduling. One can observe from this figure that the average SE achieved by the proposed CIR prediction and scheduling approach is higher than those of random scheduling and outdated CIR ones, and its performance is closer to the average SE achieved by the MMSE predictor approach which utilizes the CIR statistics. The best average SE is achieved when the CIR is known perfectly which is expected.

---

[5] Note that similar average performance is observed for other sub-carriers.
[6] Note that the proposed RLS predictor cannot perform better than the MMSE approach as the latter assumes perfect CIR statistical knowledge.

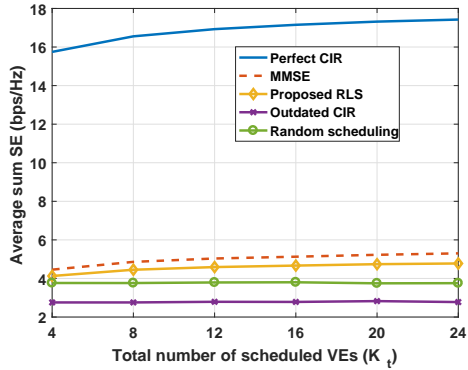

Fig. 4. Effect of the total number of scheduled VEs $K_t$.

Next we examine the effect of $K_{max}$ on the performance of the approaches when $\gamma$ is set to 10 dB as shown in Fig. 3. One can observe from this figure that the sum SE degrades with $K_{max}$ for all approaches except the perfect CIR case. The potential reason for this is that when $K_{max} > 1$, the scheduling algorithm likely supports more than one VEs and each VE will experience undesired interference which occurs due to the channel prediction error and worsens with $K_{max}$ [7]. Lastly, we evaluate the effect of $K_t$ on the performance of the approaches when $S_{max} = 2$ and $\gamma = 10$ dB as shown in Fig. 4. As can be seen from this figure, the average sum SE achieved by all approaches increase with $K_t$ up to a certain limit which fits with the phenomena observed in [13].

## V. Conclusions

This paper proposes the joint design of adaptive channel prediction, beamforming and scheduling for 5G V2I network. The channel prediction algorithm is designed without assuming the training and CIR model. In this regard, first we utilize the well known RLS technique, which does not require the channel statistics, for predicting the next block CIR. Then, we jointly design the beamforming and sub-carrier scheduling for the uplink communication to maximize the average sum rate by utilizing the predicted CIR. The receive beamforming problem is formulated as a Rayleigh quotient problem where its global optimal solution is guaranteed. And, the VE scheduling design problem is formulated as an integer programming problem where we propose the greedy based algorithm. Simulation results have confirmed the superiority of the proposed channel prediction and scheduling algorithm over the existing ones in terms of CIR accuracy and average rate.

---

[7]Recall that the proposed greedy algorithm schedules the VEs based on the predicted CIRs which may be different from the true CIRs.